\documentclass[11pt]{article}
\usepackage{moriond,epsfig}

\newcommand{\ket}[1]{\mbox{$|\!#1\;\!\rangle$}}

\def\ua{\uparrow}
\def\da{\downarrow}

\def\be{\begin{equation}}
\def\ee{\end{equation}}
\def\bea{\begin{eqnarray}}
\def\eea{\end{eqnarray}}

\def\ua{\uparrow}
\def\da{\downarrow}
\def\tp{T_{+}}
\def\tm{T_{-}}
\def\t0{T_{0}}

\begin{document}
\title{Determination of the tunnel rates through a few-electron quantum dot}

\author{\underline{R. Hanson}$^1$, I.T. Vink$^1$, D.P. DiVincenzo$^{2}$, L.M.K. Vandersypen$^1$,\\ J.M. Elzerman$^1$, L.H. Willems van Beveren$^1$ and L.P. Kouwenhoven$^1$}

\address{$^1$Kavli Institute of NanoScience Delft and ERATO Mesoscopic
Correlation Project,\\ Delft University of Technology, P.O. Box 5046, 2600 GA Delft, The Netherlands} 
\address{$^2$IBM T. J. Watson Research Center, P.O. Box 218, Yorktown Heights, NY 10598, USA}

\maketitle\abstracts{We demonstrate how rate equations can be employed to find analytical expressions for the sequential tunneling current through a quantum dot as a function of the tunnel rates, for an arbitrary number of states involved. We apply this method at the one-to-two electron transition where the electron states are known exactly. By comparing the obtained expressions to experimental data, the tunnel rates for six transitions are extracted. We find that these rates depend strongly on the spin and orbital states involved in the tunnel process.}

\section{Introduction}
Quantum dots in semiconductor heterostructures have proven to be excellent systems to study the charge and spin degree of freedom of single electrons~\cite{LeoFewEl} as well as coherent many-body processes like the Kondo effect~\cite{LeoKondo}. Quantum dots that are defined electrostatically in a two-dimensional electron gas are especially interesting, since here both the electron number and the tunnel coupling to the leads are fully tunable. The experiments in these systems are currently at the level of studying and manipulating the coherence properties of single electrons. Besides being interesting from a fundamental point of view, coherent control is an essential element of proposals to use the spin of a single electron as a building block for a quantum computer~\cite{LossDiVincenzo}.

In order to benefit fully from the tunability of the tunnel barriers, one needs to know the tunnel rate for all possible charge transitions. It was recently demonstrated that the tunnel rate through a single barrier can be determined by applying voltage pulses to one of the surface gates and monitoring the response of the quantum dot with a charge detector~\cite{JeroAPL}. This method can also give the rates of transitions involving excited states, but, due to the bandwidth of most measurement setups, its applicability lies mainly in the regime of very weak coupling to the leads (tunnel rate $<$ 100 kHz). For stronger coupling, the tunnel rates can in principle be determined from the measured current as a function of the source-drain bias voltage, by solving rate equations. This has been demonstrated for a system with two allowed transitions, both in the limit of very asymmetric and of completely symmetric barriers \cite{FujisawaPhysB,Bonet}.

In this work, we demonstrate that an expression for the current through the dot as a function of the tunnel rates can be found for an arbitrary number of possible transitions using rate equations. Then, we apply this method at the
1$\leftrightarrow$2 electron transition and give exact expressions for the case of symmetric barriers. By comparing these expressions to measurements, we determine the tunnel rates of the six lowest-energy transitions. Finally, we comment on the observed dependence of the tunnel rate on the orbital and spin state involved in the tunnel process.

\section{Calculation of the current using rate equations}

\subsection{Model and assumptions}
The system we study is a quantum dot weakly coupled to a source and a drain reservoir. We calculate the sequential tunneling current flowing through the dot, thus neglecting higher-order tunneling events. Also, we assume that relaxation processes within the quantum dot are much slower than the tunnel rates. This is generally true for relaxation between states with different spin (which we will consider in this work), but not necessarily for relaxation between states with the same spin \cite{FujisawaNature,ZeemanPRL}.

Since in our experiments both the energy spacing of the quantum dot states and the source-drain bias are much larger than the electron temperature, we take $T$ = 0. Thus, all states below the electrochemical potentials of the reservoirs are filled and all states above are empty. This implies that when a bias voltage is applied between the source and the drain, the electrons can only tunnel \textit{onto} the dot from the source contact and can only tunnel \textit{off} the dot to the drain, i.e. they tunnel in the forward direction only (exceptions will be explicitly stated).

We focus on the case where the tunnel rates through the source and the drain barrier are equal, since this simplifies the resulting equations considerably. In the experiments, the barriers can easily be tuned to make this assumption valid.
\subsection{Calculation of steady-state occupation probabilities}
To find the current flowing through the dot, we first calculate the occupation probabilities for all relevant states by solving the rate equations for the available transitions. Consider a dot where \textit{N}+1 energy states contribute to charge transport. The time evolution of the occupation probability $\rho_0$ of an energy state $|0\rangle$ depends on the occupation probabilities of the other energy states and the tunnel rates between the states:
\begin{equation}\label{RateEqn}
	\dot{\rho}_{0}=\frac{\partial\rho_{0}}{\partial t}=-\sum_{i=1}^{N}\Gamma_{i,0}\rho_{0}+\Gamma_{0,1}\rho_{1}+\cdots+\Gamma_{0,N}\rho_{N},
\end{equation}
where $\Gamma_{f,i}$ is defined as the tunnel rate between the initial state $|i\rangle$ and final state $|f\rangle$. We can write down the rate equations for all \textit{N}+1 levels conveniently within a matrix representation:
\begin{equation}\label{Matrix equation}
\frac{\partial}{\partial t}\left(
\begin{array}{c}
\rho_{0}\\
\vdots\\
\rho_{N}\\
\end{array}\right)=\left(
\begin{array}{cccc}
-\sum_{i=1}^{N}\Gamma_{i,0} & \Gamma_{0,1} & \ldots & \Gamma_{0,N} \\
\vdots & \vdots & & \vdots \\
\Gamma_{N,0} & \Gamma_{N,1} & \ldots & -\sum_{i=0}^{N-1}\Gamma_{i,N} \\
\end{array}\right) \left(
\begin{array}{c}
\rho_{0}\\
\vdots\\
\rho_{N}\\
\end{array} \right)= \mathbf{\Gamma}\left(
\begin{array}{c}
\rho_{0}\\
\vdots\\
\rho_{N}\\
\end{array}\right).
\end{equation}
We call $\mathbf{\Gamma}$ the transition matrix. The first column of this matrix (except for the diagonal element) consists of the probabilities per unit time, $\Gamma_{k,0}$, that the dot makes a transition to state $|k\rangle$ if it was initially in $|0\rangle$. These probabilities follow Poissonian statistics. The next $N$ columns correspond to the dot initially being in $|1\rangle$, $\ldots$, $|N\rangle$, respectively. The diagonal elements of the matrix, $\Gamma_{ii}$ depend on the non-diagonal elements via the relation, $\Gamma_{ii}=-\sum_{k\neq i}\Gamma_{ki}$.

We are interested in the steady-state solution of Eq. \ref{Matrix equation}. Since in this case the occupation probability distribution is constant in time, we find the steady-state occupation probability distribution by solving
\begin{equation}\label{Matrix equation equilibrium}
\frac{\partial}{\partial t}\left(
\begin{array}{c}
\rho_{0}\\
\vdots\\
\rho_{N}\\
\end{array}\right)=\mathbf{\Gamma}\left(
\begin{array}{c}
\rho_{0}\\
\vdots\\
\rho_{N}\\
\end{array}\right)=0
\end{equation}
i.e. determine the eigenvector of $\mathbf{\Gamma}$ with
eigenvalue 0.

\subsection{Calculation of the current from the occupation probabilities}
The current can be calculated from the steady-state occupation probabilities by multiplying these probabilities with the tunnel rates through one of the barriers. The currents through the two barriers are equal in the steady state. Consider transitions between states $|i_M\rangle$ with \textit{M} electrons on the dot, having occupation probabilities $\rho_{i(M)}$, and states $|j_{M+1}\rangle$ with \textit{M}+1 electrons on the dot, having occupation probabilities $\rho_{j(M+1)}$. If only forward tunneling is possible, the current through one barrier will be due to electrons added to the dot when it holds \textit{M} electrons, whereas the current through the other barrier is carried by electrons that are extracted from the dot when it is in one of the (\textit{M}+1)-electron states. The current is therefore

\begin{equation}\label{current}
I = e \sum_{i_M}\left(
\rho_{i(M)}\sum_{j_{M+1}}\Gamma_{j(M+1),i(M)}\right)=e \sum_{j_{M+1}}\left(
\rho_{j(M+1)}\sum_{i_{M}}\Gamma_{i(M),j(M+1)}\right)
\end{equation}
If electrons can also tunnel in the opposite direction (reverse tunneling), the current can still be calculated in a similar way by making suitable substitutions, as shown below.

\section{Expressions for the current at the 1$\leftrightarrow$2 electron transition}
\subsection{Energies and electrochemical potentials of the 1$\leftrightarrow$2 electron transition}
We apply the method explained in Section 2 to the transition between 1 and 2 electrons on the dot and find expressions for the current for different source-drain bias conditions. 
First we write down the energies of the relevant states and from these we deduce the electrochemical potentials of all the allowed transitions. To observe spin effects, we assume that a large in-plane magnetic field $B_{/\!/}$ is applied to the dot, which has a negligible effect on the orbitals but causes a large Zeeman energy splitting $\Delta E_{Z}\!= g \mu_B B_{/\!/}$ between the spin states \cite{ZeemanPRL}. The two lowest-energy one-electron states are the two spin states of the lowest orbital, \ket{\ua} and \ket{\da}, with energies $E_{\ua}$ and $E_{\da}\!=\!E_{\ua}\!+\!\Delta E_{Z}$ respectively.

The ground state for two electrons on the dot is a spin singlet (total spin quantum number $S\!=\!$ 0) \cite{Ashcroft}, formed by the two electrons occupying the lowest orbital with their spins anti-parallel: $\ket{\:S}\!=\!(\ket{\ua\da}\!-\!\ket{\da\ua})/\sqrt{2}$. The first excited states are the spin triplets ($S\!=\!$ 1), where the antisymmetry of the two-electron wave function requires one electron to occupy a higher orbital. The three triplet states are degenerate at zero magnetic field, but acquire different Zeeman energy shifts in finite magnetic fields because their spin $z$-components (quantum number $m_S$) differ: $m_S\!=\!+1$ for $\ket{\:T_{+\!}}\!=\!\ket{\ua\ua}$, $m_S\!=\!0$ for $\ket{\:T_{0}}\!=\!(\ket{\ua\da}\!+\!\ket{\da\ua})/\sqrt{2}$ and $m_S\!=\!-1$ for $\ket{\:T_{-\!}}\!=\!\ket{\da\da}$. We can write the energies of the two-electron states in terms of the single-particle energies of the two electrons plus a charging energy $E_{C}$ which accounts for the Coulomb interactions:
\begin{eqnarray*}
	&&\!\!\!\!\!\!\!\!E_S\ =\! E_{\ua}+ E_{\da} + E_{C} = 2 E_{\ua}+ \Delta E_{Z} + E_{C}\\
	&&\!\!\!\!\!\!\!\!E_{T_+}\!=\! 2 E_{\ua} + E_{ST} + \!E_{C}\\
	&&\!\!\!\!\!\!\!\!E_{T_0} =\! E_{\ua}\! + \! E_{\da}\! + \! E_{ST}\!+\!E_{C} = 2 E_{\ua}\!+\! E_{ST}\! +\! \Delta E_{Z}\! + \! E_{C}\\
	&&\!\!\!\!\!\!\!\!E_{T_-}\! =\! 2E_{\da}\!+\!E_{ST}\!+\!E_{C}=2 E_{\ua}\!+\!E_{ST}\!+\! 2 \Delta E_{Z} \!+\!E_{C},
\end{eqnarray*}
with $E_{ST}$ denoting the singlet-triplet energy difference in the absence of Zeeman splitting.

The electrochemical potentials $\mu$ for the different transitions are now easily calculated using $\mu_{a \leftrightarrow b}=E_b-E_a~$ \cite{LeoFewEl}. This yields for the 1$\leftrightarrow$2 electron transitions:
\begin{eqnarray}
	\mu_{\ua \leftrightarrow S}&=& E_{\ua} +E_{C} + \Delta E_Z \\
	\mu_{\da \leftrightarrow S}&=& E_{\ua} + E_{C}\\
	\mu_{\ua \leftrightarrow T_+} =\, \mu_{\da \leftrightarrow T_0} &=& E_{\ua} +E_{C} + E_{ST} \\
	\mu_{\ua \leftrightarrow T_0} =\, \mu_{\da \leftrightarrow T_-}\! &=& E_{\ua} +E_{C} + \Delta E_Z  + E_{ST}.
\end{eqnarray}
We have omitted the transitions $\ua \leftrightarrow \!T_-$ and $\da \leftrightarrow \!T_+$, since these transitions require a change in the spin $z$-component of more than $\frac{1}{2}$ and thus a spin-flip is needed in the tunneling process. Since the single-spin Zeeman relaxation \cite{ZeemanPRL} as well as the triplet-to-singlet relaxation \cite{FujisawaNature} is very slow, we can neglect these spin-blocked transitions.

\subsection{Solving the rate equations of the 1$\leftrightarrow$2 electron transition}
\begin{figure}[!h]
  \centering
  \includegraphics[width=4.5in]{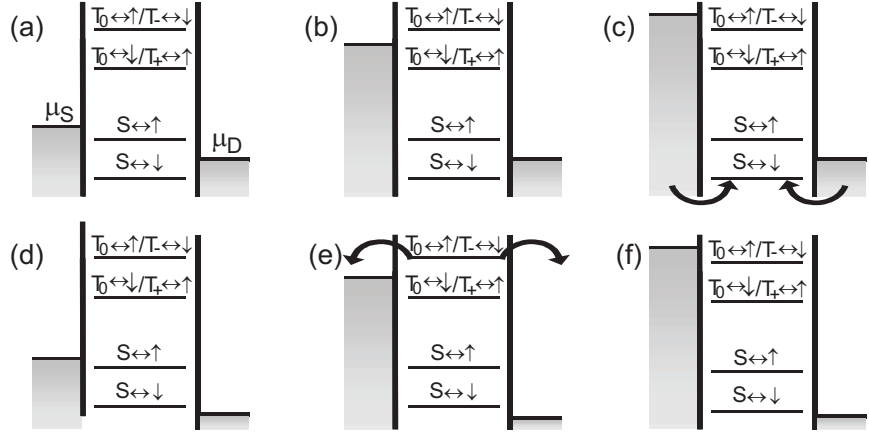}
  \caption{(a)-(f) Electrochemical potential diagrams at the 1$\leftrightarrow$2 electron transition, corresponding to the configurations $A$-$F$ respectively. Some transitions (indicated by arrows) are possible even when the corresponding electrochemical potential is not in the bias window.}
  \label{FilterEnergyDiagrams}
\end{figure}

By adjusting the electrochemical potentials of the source, $\mu_S$, and the drain, $\mu_D$, relative to the ladder of electrochemical potentials in the dot, six different configurations can be obtained where current flows. These configurations are depicted in Figs. \ref{FilterEnergyDiagrams}a-f, and we label them $A$ through $F$, respectively. (Note that current can \textit{only} flow if the transition between the one-electron and two-electron ground states, $\ua \leftrightarrow\! S$, is in the bias window, i.e. $\mu_S>\mu_{\ua \leftrightarrow S}>\mu_D$.)

To obtain expressions for the current, we first identify which transitions are possible for a specific configuration and set the tunnel rate for all other transitions to zero. Also, we take into account that for some configurations electrons can tunnel in both forward and reverse direction, which yields a doubling of the transition probability. Then, we find the steady-state occupation probabilities of all the states in terms of the tunnel rates by calculating the eigenvector of the resulting transition matrix with eigenvalue zero. This is done using the program \textit{Mathematica}. From the occupation probabilities we obtain an analytical expression for the current as a function of the tunnel rates, as explained before. We now demonstrate the construction of the transition matrix for configurations $C$ and $F$.

In the basis $\rho=(\rho_{\ua},\rho_{\da},\rho_{S},\rho_{\tp},\rho_{\t0},\rho_{\tm})$, the full transition matrix for the 1$\leftrightarrow$2 electron transition is
\begin{eqnarray}\label{TransMatrix}
\!\mathbf{\Gamma}\!=\!\!\left(\begin{array}{cccccc}
-\!\Gamma_{S\uparrow}\!-\!\Gamma_{T_{+}\uparrow}\!-\!\Gamma_{T_{0}\uparrow}&0& \Gamma_{\uparrow S}&\Gamma_{\uparrow T_{+}}&\Gamma_{\uparrow T_{0}}&0 \\
0 & -\!\Gamma_{S\downarrow}\!-\!\Gamma_{T_{0}\downarrow}\!-\!\Gamma_{T_{-}\downarrow} & \Gamma_{\downarrow S} & 0 & \Gamma_{\downarrow T_{0}} & \Gamma_{\downarrow T_{-}} \\
\Gamma_{S\uparrow} & \Gamma_{S\downarrow} & -\!\Gamma_{\uparrow S}\!-\!\Gamma_{\downarrow S} & 0 & 0 & 0 \\
\Gamma_{T_{+}\uparrow} & 0 & 0 & -\!\Gamma_{\uparrow T_{+}} &
0 & 0 \\
\Gamma_{T_{0}\uparrow} & \Gamma_{T_{0}\downarrow} & 0 & 0 &
-\!\Gamma_{\uparrow T_{0}}\!-\!\Gamma_{\downarrow T_{0}} & 0 \\
0 & \Gamma_{T_{-}\downarrow} & 0 & 0 & 0 &
-\!\Gamma_{\downarrow T_{-}}
\end{array} \right)&&
\end{eqnarray}
In configuration $F$, all transitions are possible and electrons can only flow in one direction (see Fig. \ref{FilterEnergyDiagrams}f), and therefore Eq. \ref{TransMatrix} directly gives us the relevant matrix.

In configuration $C$, the transition from $|S\rangle$ to \ket{\da} is blocked, and the transition from \ket{\da} to $|S\rangle$ is possible via both barriers. Thus, we get the transition matrix for $C$ by making the substitutions $\Gamma_{\downarrow S}\rightarrow0$ and $\Gamma_{S \downarrow}\rightarrow 2 \Gamma_{S \downarrow}$ in Eq. \ref{TransMatrix}.

The transition matrices for the other configurations can be derived in a similar way. The calculated expressions for the current in configurations $A$ through $F$ are given in the Appendix.

\section{Extraction of the tunnel rates from experimental data}
\begin{figure}[h]\label{Fig2}
\centering
\psfig{figure=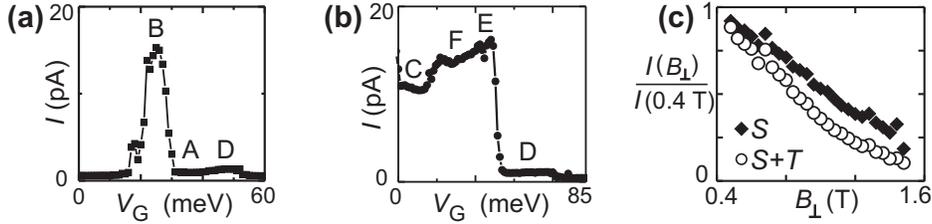,height=1.2in}
\caption{(a)-(b) Current $I$ as a function of $V_G$ at $B_{/\!/}$ = 12 T for (a) $V_{SD}$ = 310 $\mu$V and (b) $V_{SD}$ = 800 $\mu$V. In these graphs, $A$-$F$ indicate the configuration of the levels. (c) The current $I$ via the singlet state (S) and via both the singlet and all triplet states (S+T) as a function of $B_{\perp}$ ($V_{SD}$=1 meV). The currents are normalized to the values at $B_{\perp}$=0.4 T. The Zeeman energy is not resolved at these low magnetic fields.}
\end{figure}

In this section, we use the expressions for the current obtained in the previous section (and given in the Appendix) to extract the tunnel rates for all six transitions from experimental data at $B_{/\!/}$ = 12 T. This data was previously analyzed in the context of spin filtering. Experimental details can be found in Ref. \cite{SpinFilter}. In Figs. \ref{Fig2}a and \ref{Fig2}b we present traces of the current through the dot as a function of gate voltage $V_G$. Changing $V_G$ shifts the whole ladder of electrochemical potentials and thus allows us to scan through different configurations as indicated by letters $A$ through $F$. The two traces in Figs. \ref{Fig2}a and \ref{Fig2}b are taken at different source-drain bias voltage $V_{SD}$. Due to capacitive coupling of the reservoirs to the dot, changing $V_{SD}$ also slightly changes the barriers and thereby the absolute values of the tunnel rates. However, we assume the \textit{relative} values of the tunnel rates to be independent of $V_{SD}$ and thus the ratio \textit{between} the rates to be constant.

From the trace shown in Fig. \ref{Fig2}a we find that the current in configuration $A$ is 0.41~pA, yielding the tunnel rate $\Gamma_{S\ua}$~=~5.1~MHz. The measured current in configuration $D$ is 0.77~pA, which leads to  $\Gamma_{S\da}$ = 9.3~MHz, using $\Gamma_{S\ua}$~=~5.1~MHz. We note that there is a spin dependence in the tunnel rate: $\Gamma_{S\da}\approx 2 \Gamma_{S\ua}$. This spin dependence has been observed before at the 0$\leftrightarrow$1 electron transition in the same device \cite{SpinFilter}, and might be attributed to exchange interactions in the leads close to the dot. The ratio for spin-up and spin-down electrons tunneling to the same orbital, $\alpha$, is assumed to be constant around the $1\!\leftrightarrow\!2$ electron transition and independent of $V_{SD}$ and $V_{G}$. In the rest of this section we use $\alpha$ = 2.

The current in region $A$ is 0.44 pA close to region $B$, and the tunnel rate $\Gamma_{S\ua}$, is calculated to be 5.5~MHz. It differs slightly from the value close to region $D$ obtained above because the barriers are not completely independent of $V_G$. In region $B$, the measured current is 14~pA, yielding $\Gamma_{\tp\ua}$ = 0.27~GHz.

We now turn our attention to Fig. \ref{Fig2}b. Here, the current in configuration $D$ is $I_{D}$ = 0.60~pA, which gives $\Gamma_{S\ua}$ = 3.8~MHz, $\Gamma_{S\da}$ = 7.6~MHz and $\Gamma_{\tp\ua}$ = 0.18~GHz, assuming the ratio of the rates is fixed. For the current in configuration $E$ we find $I_{E}$ = 13~pA. From this number we deduce $\Gamma_{\t0\ua}$ = 0.09~GHz and $\Gamma_{\t0\da}=\alpha \Gamma_{\t0\ua}$ = 0.18~GHz. From the magnitude of the current in $F$, $I_{F}$ = 14~pA, we find $\Gamma_{\tm\da}$ = 0.18~GHz.

Since all the rates have been determined, we can predict the current in configuration $C$. The rates determined above predict a current of 9.2~pA, in good agreement with the measured value of 10~pA.

We finally comment on the strong dependence of the tunnel rates on the orbital state involved in the tunnel process: tunneling to a triplet state is more than an order of magnitude faster than tunneling to the singlet state. This effect can be explained by the different spatial distribution of the singlet state and the triplet states. In a singlet state, both electrons are in the lowest orbital state, and are therefore mainly located near the center of the dot \cite{LeoFewEl}. In a triplet state, one electron occupies an orbital excited state which has more weight near the edge of the dot, and therefore is expected to have a larger overlap with the reservoirs. To illustrate the significance of the spatial weight, results are presented from a measurement on a similar device in a perpendicular magnetic field $B_{\perp}$. The current via the singlet state and the current via both the singlet and the triplets are plotted as a function of $B_{\perp}$ in Fig. \ref{Fig2}c. The perpendicular field causes the wave functions to shrink in the plane of the two-dimensional electron gas, reducing the overlap with the leads and thereby reducing the corresponding tunnel rates. The decrease in current for increasing $B_{\perp}$ is clearly observed.

\section*{Acknowledgments}
This work was supported by the DARPA-QUIST program, the ONR and the EU-RTN network on spintronics. 

\section*{Appendix}
The analytical expressions for the current as a function of the tunnel rates in configurations $A$ through $F$, $I_A$ through $I_F$, are:
\begin{eqnarray}
I_{A}&=& \frac{1}{2}e\Gamma_{S\uparrow} \\
I_{B}&=& \frac{1}{3}e\left(\Gamma_{S\uparrow}+\Gamma_{T_{+}\uparrow}\right) \\
I_{C} &=& \frac{e}{6+\Gamma_{\downarrow
S}\left(\frac{8}{\Gamma_{\downarrow
T_{0}}}+\frac{2}{\Gamma_{\uparrow
S}}+\frac{6}{\Gamma_{\uparrow T_{0}}}\right)} \times \\
& & \left(\left(1+2\frac{\Gamma_{\downarrow S}}{\Gamma_{\downarrow
T_{0}}}+2\frac{\Gamma_{\downarrow S}}{\Gamma_{\uparrow
T_{0}}}\right)\left(\Gamma_{S\uparrow}+\Gamma_{T_{+}\uparrow} +
\Gamma_{T_{0}\uparrow}\right) + \left(
\Gamma_{S\downarrow}+\Gamma_{T_{-}\downarrow}+\Gamma_{T_{0}\downarrow}
\right) \right) \nonumber \\
I_{D} &=& \frac{1}{3}e(\Gamma_{S\uparrow}+\Gamma_{S\downarrow}) \\
I_{E} &=& \frac{e}{5+\left(\frac{6\Gamma_{\uparrow
T_{0}}}{\Gamma_{\downarrow S}}+\frac{8\Gamma_{\uparrow
T_{0}}}{\Gamma_{\downarrow T_{0}}}+\frac{4\Gamma_{\uparrow
T_{0}}}{\Gamma_{\uparrow S}}\right)} \times \\
& & \left(\frac{\Gamma_{\da S}\Gamma_{\da\t0}\Gamma_{\ua
S}+2\Gamma_{\da S}\Gamma_{\da\t0}\Gamma_{\ua\t0}+2\Gamma_{\da
S}\Gamma_{\ua S}\Gamma_{\ua\t0}+2\Gamma_{\da\t0}\Gamma_{\ua
S}\Gamma_{\ua\t0}}{\Gamma_{\da S}\Gamma_{\da\t0}\Gamma_{\ua
S}}\right)\times \nonumber \\
& & \left(\left(\Gamma_{S\uparrow}+\Gamma_{T_{+}\uparrow} +
\Gamma_{T_{0}\uparrow}\right) +
\left(\frac{\Gamma_{\da\t0}+2\Gamma_{\ua\t0}}{\Gamma_{\da\t0}}\right)\left(
\Gamma_{S\downarrow}+\Gamma_{T_{-}\downarrow}+\Gamma_{T_{0}\downarrow}
\right)\right) \nonumber \\
I_{F} &=& \frac{1}{6}e(\Gamma_{S\uparrow}+\Gamma_{T_{+}\uparrow}+\Gamma_{T_{0}\uparrow}+\Gamma_{S\downarrow}+\Gamma_{T_{-}\downarrow}+\Gamma_{T_{0}\downarrow})
\end{eqnarray}

\end{document}